\documentclass[prl,reprint,amsmath,amssymb,aps,preprintnumbers,showpacs,twocolumn]{revtex4}
\usepackage[colorlinks=true,linkcolor=black, citecolor=black,urlcolor=black]{hyperref} 
\usepackage{graphicx}  
\usepackage{braket} 
\usepackage{bbm} 
\usepackage{tikz,tikz-cd} 
\usepackage[utf8]{inputenc}

\usepackage{comment}

\usetikzlibrary{calc}
\usetikzlibrary{arrows}

\def\be{\begin{equation}}
\def\ee{\end{equation}}

\DeclareMathOperator{\tr}{tr}

\newcommand{\AD}{\operatorname{Ad}}

\begin{document}

\vspace*{0cm}

\title{Quantum correction to generalized T-dualities}
\author{Riccardo Borsato}
\email{riccardo.borsato@usc.es}
\affiliation{Instituto Galego de F\'isica de Altas Enerx\'ias (IGFAE), Universidade de  Santiago de Compostela, Spain}
\author{Linus Wulff}
\email{wulff@physics.muni.cz}
\affiliation{Department of Theoretical Physics and Astrophysics, Masaryk University, Brno, Czech Republic}


\begin{abstract}
Poisson-Lie duality is a generalization of abelian and non-abelian T-duality, and it can be viewed as a map between solutions of the low-energy effective equations of string theory, i.e. at the (super)gravity level. We show that this fact extends to the next order in $\alpha'$ (two loops in $\sigma$-model perturbation theory) provided that the map is corrected. The $\alpha'$-correction to the map is induced by the anomalous Lorentz transformations of the fields that are necessary to go from a doubled $O(D,D)$-covariant formulation to the usual (super)gravity description.
\end{abstract}


\maketitle


\textbf{Introduction.}
The notion of T-duality \cite{Buscher:1987sk,Buscher:1987qj} is central in string theory. It says that a closed string on a background with abelian isometries has another description as a string on a dual background. In the simplest case of T-duality on a circle, the duality acts by inverting the radius of the circle. More generally, backgrounds may have non-abelian isometry groups, and at least at the classical level there is indeed a generalization to a non-abelian version of T-duality \cite{delaOssa:1992vci}. Unlike in the abelian case, non-abelian T-duality (NATD) does not generically preserve the isometries of the background, and it is therefore not obvious how to invert the transformation.
This problem was overcome by Klim\v c\'ik and \v Severa in \cite{Klimcik:1995ux,Klimcik:1995dy}. They realized that the map can be made invertible by relaxing the notion of isometry. One requires the background to have instead so-called Poisson-Lie (PL) symmetry, namely to possess vector fields $v_i$, with $[v_i,v_j]=-f_{ij}{}^kv_k$, under which the metric and $B$-field of the $\sigma$-model transform as 
\begin{equation}\label{eq:PL-symm}
\mathcal L_{v_i}M_{mn}=-\tilde f^{jk}{}_i\ v_j{}^pv_k{}^q\ M_{mp}M_{qn}\,,
\end{equation}
where $M_{mn}=G_{mn}-B_{mn}$ and $\tilde f^{jk}{}_i$ are structure constants of a dual Lie algebra. This more general notion of symmetry allows to define a dual background (see below). This construction became known as ``Poisson-Lie T-duality'' since the group structure underlying it is that of a PL group. The $\sigma$-models on the original and dual backgrounds are classically equivalent being related by a canonical transformation~\cite{Sfetsos:1997pi}. When the dual structure constants $\tilde f$ vanish,  $v_i$ generate standard isometries, and one recovers (N)ATD. 

At the worldsheet quantum level, i.e. including string $\alpha'$-corrections, things are more subtle. While abelian T-duality remains a symmetry of the worldsheet CFT to all orders in $\alpha'$, it was quickly realized that NATD cannot be a symmetry at the quantum level \cite{Giveon:1993ai}. At best it can map one worldsheet CFT to another --- inequivalent --- one. It can therefore be used to generate new string backgrounds from old ones. Except for an anomaly when dualizing non-unimodular groups \cite{Alvarez:1994np,Elitzur:1994ri}, this has been shown to work to zeroth order in $\alpha'$, i.e. at the low-energy (supergravity) level of the string effective equations, which corresponds to one loop order in $\sigma$-model perturbation theory. Similar results are known for PL duality, see e.g.~\cite{Alekseev:1995ym,Tyurin:1995bu}. It has been a long standing problem whether PL and NATD can be extended beyond this lowest order.

Here we show that PL duality can be extended to order $\alpha'$, i.e. two loops in the $\sigma$-model perturbation theory, provided that the map is corrected. A special case of our results gives the corrections to NATD. When specifying to the abelian case we recover the results of~\cite{Kaloper:1997ux}.

To find this correction we exploit a powerful formulation of the string effective equations inspired from Double Field Theory (DFT). It has long been known that the bosonic string compactified on a $d$-torus has an $O(d,d)$ T-duality symmetry~\cite{Giveon:1994fu}. DFT is a field theory where this symmetry is made manifest form the start~\cite{Siegel:1993xq,Siegel:1993th,Aldazabal:2013sca,Berman:2013eva,Hohm:2013bwa} and is therefore well suited to working with T-duality. This is achieved by doubling the dimension of the physical manifold, and by imposing a ``section condition'' which effectively eliminates the dependence on half of the coordinates, giving the correct dimension in the end ($D=26$ for the bosonic string and $D=10$ for the superstring). Here we always work with the standard choice of section, so that the background depends only on the physical coordinates. In this formulation it is rather the dimension of the \emph{tangent} space that is doubled, and we have two copies of the Lorentz group instead of one~\cite{Hohm:2010xe}. The standard Lorentz group is the diagonal of the doubled one, and under this breaking the equations of DFT reduce to the standard string effective equations, at lowest order in $\alpha'$. A crucial point is that at the quantum level it is impossible to preserve both the $O(D,D)$ and the Lorentz covariance of the fields \cite{Hohm:2011si,Hohm:2013jaa,Hohm:2014xsa,Hohm:2014eba}. If we insist on fields which transform nicely under T-duality and $O(D,D)$,  they must transform non-covariantly under Lorentz transformations~\cite{Marques:2015vua} (see \cite{Eloy:2019hnl} for another manifestation of this fact). The fact that the Lorentz transformation of the fields receives corrections at order $\alpha'$ makes the discussion of the Lorentz invariance of the theory non-trivial. But this can be turned into a virtue rather than a shortcoming. In fact the $\alpha'$-correction to the Lorentz transformation fixes the correction to the DFT action \cite{Marques:2015vua}. Remarkably, this $\alpha'$-corrected $O(D,D)$-covariant action correctly reproduces the $\alpha'$-corrections to the bosonic and heterotic string effective actions~\cite{Marques:2015vua}.

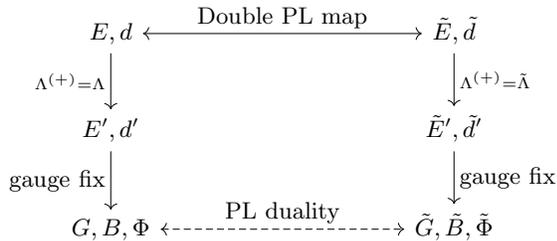
\begin{figure}%
\[
\begin{tikzcd}[column sep=10em, row sep=2.2em]
E, d\arrow[swap]{d}{\Lambda^{(+)}=\Lambda} \arrow[leftrightarrow]{r}{\mbox{Double PL map}} & \tilde E,\tilde d\arrow{d}{\Lambda^{(+)}=\tilde\Lambda}\\
E', d' \arrow[swap]{d}{\mbox{gauge fix}} & \tilde E',\tilde d' \arrow{d}{\mbox{gauge fix}}\\
G,B,\Phi \arrow[dashed,leftrightarrow]{r}{\mbox{PL duality}} & \tilde G,\tilde B,\tilde\Phi 
\end{tikzcd}
\]
\caption{Starting with the PL duality map for the doubled fields $(E,d)$, the map for the standard (super)gravity fields $(G,B,\Phi)$ is obtained after a double Lorentz transformation $(\Lambda^{(+)},\Lambda^{(-)})=(\Lambda,1)$ to set $e^{(+)}=e^{(-)}$, thus breaking the double Lorentz group down to its diagonal subgroup. The $\alpha'$-corrections to the PL duality map follow from the anomalous Lorentz transformations of the fields.}%
\label{fig:PL-diagram}%
\end{figure}

Our strategy is to use the rewriting of PL duality in the doubled language, where it takes a natural form, see e.g. \cite{Hassler:2017yza,Jurco:2017gii,Demulder:2018lmj,Sakatani:2019jgu,Catal-Ozer:2019hxw}. The basic fields of the formulation we use, the ``generalized fluxes'', turn out to be invariant under PL duality. Since the string effective equations, including the first $\alpha'$-correction, can be written in terms of the generalized fluxes~\cite{Marques:2015vua,Baron:2017dvb}, at least to this order PL duality maps solutions of the doubled equations to solutions. 
At the standard (super)gravity level there are in fact explicit corrections to the PL duality rules. They arise from the non-covariance of the doubled fields under the double Lorentz transformation needed to gauge-fix down to the diagonal subgroup and to go to the standard (non-doubled) description. See Figure \ref{fig:PL-diagram} for a summary. This strategy was used in \cite{Borsato:2020bqo} to find the $\alpha'$-correction to the ``homogeneous Yang-Baxter deformations'' (related to NATD \cite{Hoare:2016wsk,Borsato:2016pas})
and it works for any $O(D,D)$ transformation leaving the generalized fluxes invariant.

\textbf{Poisson-Lie duality.}
In PL duality,  $f_{ij}{}^k$ and $\tilde f^{ij}{}_k$ are interpreted as structure constants of Lie groups denoted by $G$ and $\tilde G$. These  are combined into a ``Drinfel'd double'' $\mathcal D$ whose Lie algebra is generated by $T_I=(T_i,\tilde T^i)$, where $T_i$ are generators of Lie$(G)$, and $\tilde T^i$ of Lie$(\tilde G)$. Obviously Lie$(G)$ and Lie$(\tilde G)$ are subalgebras of $\mathcal D$ but there are also mixed commutation relations
\begin{equation}
\begin{aligned}
&[T_i,T_j]=f_{ij}{}^kT_k\,,\qquad
[\tilde T^i,\tilde T^j]=\tilde f^{ij}{}_k\tilde T^k\,,\\
&[T_i,\tilde T^j]=\tilde f^{jk}{}_iT_k-f_{ik}{}^j\tilde T^k\,.
\end{aligned}
\end{equation}
Importantly, $\mathcal D$ is endowed with the invariant symmetric bilinear form $\braket{T_I,T_J}$ defined by
\begin{equation}
\braket{T_i,T_j}=\braket{\tilde T^i,\tilde T^j}=0\,,\qquad
\braket{T_i,\tilde T^j}=\delta_i^j\,.
\end{equation}

Having introduced $\mathcal D$ we can now present PL duality as an invertible map between an ``original'' background (specified by  a metric $G_{mn}$, a Kalb-Ramond field $B_{mn}$ and a dilaton $\Phi$) and another ``dual'' background (with fields $\tilde G_{mn},\tilde B_{mn}$ and $\tilde \Phi$). 
We  split the coordinates of the original background as $x^m=(y^\sigma,x^\mu)$, where $y^\sigma$ are  coordinates on the group $G$ to be dualized, and $x^\mu$ are  coordinates that play the role of spectators under the dualization. Similarly, for the dual background we  have $\tilde x^m=(\tilde y^\sigma,x^\mu)$ with $\tilde y^\sigma$ coordinates on $\tilde G$. The $y$ and $\tilde y$-dependence is in fact encoded in the group elements $g(y)\in G$ and $\tilde g(\tilde y)\in\tilde G$ featuring below.
To present the map between the original and dual backgrounds we first need the fact that the condition~\eqref{eq:PL-symm} implies that $M_{mn}\equiv G_{mn}-B_{mn}$ is of the form
\begin{equation}\label{eq:M}
M=U\dot M(1+\Pi \dot M)^{-1} U^T\,,
\end{equation}
where we suppressed matrix indices for readability. The matrix $U_m{}^r$ depends only on $y$ and it is of block form with non-vanishing components $U_\mu{}^\nu=\delta_\mu{}^\nu$ and $U_\sigma{}^i=u_\sigma{}^i$, the latter being the components of the Maurer-Cartan form $u=g^{-1}vg=g^{-1}dg=dy^\sigma u_\sigma{}^iT_i$~\footnote{We will use indices $r,s$ where $r=(i,\mu)$ to distinguish them from indices $m,n$ where $m=(\sigma,\mu)$.}. The matrix $\Pi^{rs}$ depends only on $y$ and its only non-trivial components are
\begin{equation}\label{eq:Pi}
\Pi^{ij} = \braket{\AD_g^{-1}\circ{\rm P}\circ\AD_g\tilde T^i,\tilde T^j}\,,
\end{equation}
where $\AD_gX=gXg^{-1}$ and ${\rm P}$ is the projector on Lie$(G)$. Notice that in general $\Pi\neq 0$ thanks to the mixed commutation relation of $\mathcal D$ if $\tilde f^{ij}{}_k\neq 0$.
The map between $M_{mn}$ and $\tilde M_{mn}$ is achieved by relating both of them to $\dot M_{rs}$, a matrix depending only on spectators $x^\mu$ and on which no other condition is imposed~\footnote{We obviously have to assume invertibility of the inverse matrices in~\eqref{eq:M} and~\eqref{eq:Mt}.}. The dual background $\tilde M_{mn}$ is obtained by~\footnote{See the Supplemental Material for a rewriting of this formula more familiar to the usual form of the (N)ATD rules. In the case of no spectators (and assuming invertibility of $\dot M$) Eq.~\eqref{eq:M} reduces to $M=u(\dot M^{-1}+\Pi)^{-1}u^T$ and Eq.~\eqref{eq:Mt} to $\tilde M=\tilde u(\dot M+\tilde \Pi)^{-1}\tilde u^T$, which are more symmetric and more familiar in the PL duality literature.}
\begin{equation}\label{eq:Mt}
\tilde M=\tilde U[(\dot M+\tilde \Pi)P+\bar P]^{-1}(\dot M\bar P+P)\tilde U^T\,,
\end{equation}
where $\tilde U_\mu{}^\nu=\delta_\mu{}^\nu$, $\tilde U_\sigma{}^i=\tilde u_{\sigma j}\delta^{ji}$ and $P, \bar P$ project on indices $i,j$ and $\mu,\nu$ respectively.
As previously $\tilde u=\tilde g^{-1}d\tilde g=d\tilde y^\sigma\tilde u_{\sigma i}\tilde T^i$ is a Maurer-Cartan form, and now 
$\tilde \Pi_{ij} = \braket{\AD_{\tilde g}^{-1}\circ\tilde{\rm P}\circ\AD_{\tilde g} T_i, T_j}$, where $\tilde{\rm P}$ projects on Lie$(\tilde G)$.
Finally, the dilatons of the two backgrounds are related by~\cite{VonUnge:2002xjf}
\begin{equation}\label{eq:PhiPhit}
\exp(-2\Phi)\frac{(\det G)^{1/2}}{\det u}
=
\exp(-2\tilde \Phi)\frac{(\det \tilde G)^{1/2}}{\det\tilde u}\,.
\end{equation}
Taking $\tilde G$ abelian  ($\tilde f^{ij}{}_k=0$) implies $\Pi=0$ and Eq.~\eqref{eq:M} simplifies to $M=U\dot MU^T$, encoding the usual consequences of having isometries for $M$.
 Then parameterizing the abelian group as $\tilde g=\exp(\tilde y_i \tilde T^i)$ with $\tilde y_i=\tilde y^\sigma \delta_{\sigma i}$ it follows that $\tilde u_{\sigma i}=\delta_{\sigma i}$, $\tilde \Pi_{ij}=\tilde y_kf_{ij}{}^k$, and from~\eqref{eq:Mt} and~\eqref{eq:PhiPhit} we recover the rules of NATD in the presence of spectators~\footnote{Compared to~\cite{Borsato:2018idb} here $\tilde y_i=-\nu_i$ and we swap $\tilde e^{(\pm)}$ appearing later.}.
Even simpler is the case when also $G$ is abelian ($f_{ij}{}^k=0$) so that $M$ is invariant under dim$(G)$ $U(1)$ isometries. Then also $\tilde \Pi=0$ and Eq.~\eqref{eq:Mt} implements dim$(G)$ factorized T-dualities, reducing to the celebrated Buscher rules when only one isometry is dualized.

\textbf{Double formulation.}
The non-linear maps in~\eqref{eq:M} and ~\eqref{eq:Mt} admit a much simpler and linear formulation in the doubled language, where one works with matrices $\mathcal O_M{}^N$ of dimension $2D\times 2D$. These
 are elements of the group $O(D,D)$, meaning that $\mathcal O_M{}^P\mathcal O_N{}^Q\eta_{PQ}=\eta_{MN}$ where 
\begin{equation}\label{eq:etaMN}
\eta_{MN}=
\left(
\begin{array}{cc}
0 & \delta^m{}_n\\
\delta_m{}^n & 0
\end{array}
\right)\,.
\end{equation}
In fact let us construct the (inverse) ``generalized vielbein'' which we parameterize as
\begin{equation}\label{eq:E}
E_A{}^M=
\frac{1}{\sqrt2}
\left(
\begin{array}{cc}
e^{(+)an}M_{nm} & e^{(+)am}\\
-e^{(-)}_a{}^nM_{mn} & e^{(-)}_a{}^m
\end{array}
\right)\,,
\end{equation}
where $A$ is a flat index  and  $M$ curved. We  use similar parametrizations for $\tilde E_A{}^M$ and $\dot E_A{}^R$, adding tildes and dots. Above, $e^{(\pm)}$ are two possible vielbeins for the metric $G_{mn}$. They are not necessarily equal and in general they are related by a non-trivial Lorentz transformation. Each of them transform  under only one of the two copies of the Lorentz group (distinguished by the $(+)$ and $(-)$) arising in the doubled formulation. The generalized vielbein is one of the main ingredients of the ``frame-like formulation'' of DFT, and it will be important for our derivation of the ``unimodularity condition''~\eqref{eq:sugra} and the $\alpha'$-corrections to PL duality.
It is straightforward to check that the relations~\eqref{eq:M} and~\eqref{eq:Mt} are equivalent to the relations
\begin{equation}\label{eq:EEt}
E = \dot E(1+\Pi)\mathcal U\,,\qquad
\tilde E = \dot E(1+\tilde \Pi)\tilde {\mathcal U}\,,
\end{equation}
where we suppressed indices. In our notation all dotted quantities only depend on $x^\mu$. The non-vanishing components of $\Pi_R{}^S$ and $\tilde\Pi_R{}^S$ are again only $\Pi^{ij}$ and $\tilde \Pi_{ij}$ and the antisymmetry properties $\Pi^{ij}=-\Pi^{ji}$ and $\tilde \Pi_{ij}=-\tilde \Pi_{ji}$ imply that $(1+\Pi),(1+\tilde\Pi)$ are elements of $O(D,D)$. The matrices $\mathcal U,\tilde{\mathcal U}$ are also elements of $O(D,D)$ with $\mathcal U_i{}^\sigma = u_i{}^\sigma$, $\mathcal U^i{}_\sigma = u_\sigma{}^i$, $\tilde{\mathcal U}^{i\sigma} = \tilde u^{i\sigma}$, $\tilde{\mathcal U}_{i\sigma} = \tilde{u}_{\sigma i}$, $\mathcal U_\mu{}^\nu=\mathcal U^\mu{}_\nu = \tilde{\mathcal U}_\mu{}^\nu=\tilde{\mathcal U}^\mu{}_\nu = \delta_\mu{}^\nu$.
We are using a notation so that $u_i{}^\sigma$ and $\tilde u^{i\sigma}$ are the inverses of $u_\sigma{}^i$ and $\tilde{u}_{\sigma i}$ respectively.
To match~\eqref{eq:M} and~\eqref{eq:Mt} with~\eqref{eq:EEt} one finds that the $(+)$ and $(-)$ vielbeins must transform differently
\begin{equation}
\begin{aligned}
e_a^{(\pm)m}&=\dot e_a^{(\pm)s}O_{(\pm)s}{}^r(U^{-1})_r{}^m\,,\\
\tilde e_a^{(\pm)m}&=\dot e_a^{(\pm)s}\tilde O_{(\pm)s}{}^r(\tilde U^{-1})_r{}^m\,,
\end{aligned}
\end{equation}
where 
\begin{equation}
\begin{aligned}
&O_{(+)}=1+\dot M\Pi\,,\  && \tilde O_{(+)}=\bar P+(\tilde \Pi+\dot M)P\,,\\
&O_{(-)}=1-\dot M^T\Pi,\, \  && \tilde O_{(-)}=\bar P+(\tilde \Pi-\dot M^T)P\,.
\end{aligned}
\end{equation}
In both cases the $(+)$ and $(-)$ vielbeins are then related by Lorentz transformations as $e_a^{(-)m}=\Lambda_a{}^b e_b^{(+)m}$ and $\tilde e_a^{(-)m}=\tilde \Lambda_a{}^b \tilde e_b^{(+)m}$ where
\begin{equation}\label{eq:Lor}
 \Lambda=\dot e^{-1} O_{(-)}O_{(+)}^{-1}\dot e\,,\qquad
\tilde  \Lambda=\dot e^{-1}\tilde O_{(-)}\tilde O_{(+)}^{-1}\dot e\,,
\end{equation}
if we fix $\dot e=\dot e^{(+)}=\dot e^{(-)}$.
Finally the transformation~\eqref{eq:PhiPhit} is translated into
\begin{equation}\label{eq:d}
d+\tfrac12\log\det u=\dot d=\tilde d+\tfrac12\log\det\tilde u\,,
\end{equation}
where $d,\dot d,\tilde d$ are called ``generalized dilatons'' and are parametrized as in $d=\Phi-\frac14 \log \det G$~\footnote{Because $\dot d$ depends only on $x^\mu$ it follows that $\partial_\sigma\Phi=\frac12\partial_\sigma\log\det M-\partial_\sigma\log\det u$ and similarly for $\tilde \Phi$.}.

\textbf{PL duality as a map between string backgrounds.}
The double formulation is very useful because in this language it is very simple to prove that the PL duality transformation is a solution generating technique in string theory, at least to leading and subleading order in the $\alpha'$-expansion
.
From  $E_A{}^M$ and  $d$ one can construct the ``generalized fluxes''
\begin{equation}\label{eq:Fdef}
\begin{aligned}
\mathcal F_{ABC}&=3E_{[A}{}^M\partial_M E_B{}^NE_{C]N}\,,\\
\mathcal F_{A}&=E^{BM}\partial_M E_B{}^NE_{AN}+2E_A{}^M\partial_M d\,,
\end{aligned}
\end{equation}
that are the dynamical fields of the frame-like formulation of DFT. In fact the DFT equations of motion can be written only in terms of the above fluxes and their flat derivatives $\partial_A\mathcal F=E_A{}^M\partial_M\mathcal F$, both at leading and subleading order in the $\alpha'$-expansion~\cite{Baron:2017dvb}.
Under the transformation~\eqref{eq:EEt} we have
\begin{align}
\mathcal F_{ABC}
& = 3\dot E_{[A}{}^\mu\partial_\mu \dot E_B{}^N\dot E_{C]N} \label{eq:FABC}\\
&+3\dot E_{[A}{}^i \dot E_B{}^j\dot E_{C]k}f_{ij}{}^k
+3\dot E_{[A}{}^i \dot E_{Bj}\dot E_{C]k}\tilde f^{jk}{}_i,\nonumber\\
& \nonumber\\
\mathcal F_A&=
\dot E^{B\mu}\partial_\mu \dot E_B{}^N\dot E_{AN} 
+2\dot E_A{}^\mu\partial_\mu\dot d\label{eq:FA}\\
&+\dot E_A{}^if_{ij}{}^j
-\dot E_{Ai}(\tilde f^{ij}{}_j+f_{jk}{}^{i}\Pi^{jk}).\nonumber
\end{align}
For the reader's convenience we give the details of the computation in the Supplemental Material. The results for the dual background are analogous upon exchanging tilded and untilded quantities, and appropriately raising or lowering $i,j,k$ indices. Because of the symmetry of~\eqref{eq:FABC} under this transformation, it immediately follows that $\mathcal F_{ABC}=\tilde{\mathcal F}_{ABC}$. Eq.~\eqref{eq:FA} instead is not symmetric under this transformation, but it becomes symmetric if we impose the tracelessness of the structure constants
\begin{equation}\label{eq:sugra}
f_{ij}{}^j=0\,,\qquad\qquad
\tilde f^{ij}{}_j=0\,,
\end{equation}
as detailed in the Supplemental Material. When this ``unimodularity condition'' holds we have simply
\begin{equation}
\mathcal F_A=
\dot E^{B\mu}\partial_\mu \dot E_B{}^N\dot E_{AN} 
+2\dot E_A{}^\mu\partial_\mu\dot d\,,
\end{equation}
and $\mathcal F_A=\tilde{\mathcal F}_A$ immediately follows. Notice that not only both fluxes but also their flat derivatives are invariant under the PL transformation. In fact, since they only depend on spectators $x^\mu$ it follows that $E_A{}^M\partial_M \mathcal F = 
E_A{}^\mu\partial_\mu \mathcal F = 
\tilde E_A{}^\mu\partial_\mu \tilde{\mathcal F}=
\tilde E_A{}^M\tilde\partial_M \tilde{\mathcal F}$.

If we start from a string background, or in other words given a model with $E_A{}^M$ and $d$ of the PL form~\eqref{eq:EEt} and~\eqref{eq:d} that satisfies the doubled equations of motion to zeroth and first order in $\alpha'$, we conclude that the dual model given by $\tilde E_A{}^M$ and $\tilde d$  also satisfies the same equations, at least when~\eqref{eq:sugra} holds. {This observation extends to higher orders under the assumption that there exists a formulation of the string effective action in terms of the generalized fluxes and their flat derivatives~\cite{Baron:2017dvb} also at higher orders in $\alpha'$. This in turn should be true as long as it is possible to make diffeomorphisms, B-field gauge transformations and $O(D,D)$ symmetry manifest.
}

This is a proof that PL duality is a solution generating technique in string theory at least when both structure constants are traceless, as found already to lowest order in~\cite{Bossard:2001au}. When $\tilde G$ is abelian this condition reduces to the unimodularity condition for NATD~\cite{Alvarez:1994np,Elitzur:1994ri}.

\textbf{$\alpha'$-corrections.}
So far we have shown that in the doubled formulation the PL duality transformation works and remains uncorrected at least to order $\alpha'$. 
Note that the assumption is that the DFT equations are satisfied without the need of correcting the $O(D,D)$ form~\eqref{eq:EEt} of the PL transformation, and therefore only $\dot M$ and $\dot d$ in~\eqref{eq:EEt} and~\eqref{eq:d} can depend on $\alpha'$. 

The description of the two models in terms of standard (i.e. non-doubled) fields $(G,B,\Phi)$ and $(\tilde G,\tilde B,\tilde\Phi)$ is different, and the PL duality transformation between these does receive $\alpha'$-corrections. The reason is that when going from a doubled to a standard (super)gravity formulation we must first perform a double Lorentz transformation to set the two vielbeins $e^{(+)}$ and $e^{(-)}$ equal~\cite{Marques:2015vua}.
At order $\alpha'$ the fields of the doubled formulation transform non-covariantly under local Lorentz transformations, and this induces extra $\alpha'$-corrections also for the standard fields. The situation is illustrated in figure \ref{fig:PL-diagram}.
Because of the non-covariance even under the \emph{diagonal} of the double Lorentz group, we say that the reduction from the doubled to the standard formulation picks a specific non-covariant ``scheme'', which we call the scheme of DFT. To translate our results into the covariant schemes of~\cite{Bergshoeff:1988nn,Bergshoeff:1989de,Metsaev:1987zx,Hull:1987yi} one must implement $\alpha'$-dependent field redefinitions. We provide a  dictionary~\cite{Borsato:2019oip} in the Supplemental Material.

The {first $\alpha'$-correction to} $M_{mn}$ induced by the compensating double Lorentz-transformation with  $\Lambda^{(+)}$ and $\Lambda^{(-)}$ is~\footnote{Notice the first transposition.}
\begin{equation}
a\Delta^{(-)}_{\Lambda^{(-)}} M^{(\text{DFT})}_{nm}+b\Delta^{(+)}_{\Lambda^{(+)}} M^{(\text{DFT})}_{mn}\,,
\end{equation}
where $a=b=-\alpha'$ for the bosonic string and $a=-\alpha$, $b=0$ for the heterotic string (and $a=b=0$ for type II). The finite form of the anomalous  transformations is~\footnote{We review in the Supplemental Material how to obtain the finite form of the transformation from the infinitesimal one of~\cite{Marques:2015vua}, as explained in~\cite{Borsato:2020bqo}.}
\begin{align}
\Delta^{(\pm)}_\Lambda M^{(\text{DFT})}_{mn}&=
{\tfrac12\tr\left(\partial_m\Lambda\Lambda^{-1}\ \omega^{(\pm)}_n\right)
-B^{\text{WZW},(\Lambda)}_{mn}}
\nonumber\\
&+\tfrac14\tr\left(\partial_m\Lambda \Lambda^{-1}\partial_n\Lambda\Lambda^{-1}\right)\,, \label{eq:anMpm}
\end{align}
where $\omega_{ma}^{(\pm)b}=\omega_{ma}{}^{b}\pm\frac12 H_{ma}{}^{b}$ and $\omega$ is the spin-connection for the vielbein $e$ after the diagonal gauge-fixing. The WZW-like contribution to $B$ is defined by
\be\label{eq:BWZW}
dB^{\text{WZW},(\Lambda)}=-\tfrac{1}{12}\tr\left(d\Lambda\Lambda^{-1}d\Lambda\Lambda^{-1}d\Lambda\Lambda^{-1}\right)\,.
\ee

The $\alpha'$-corrections to the original model can be obtained for example after choosing $e=e^{(-)}$ and doing the double Lorentz transformation on $e^{(\pm)}$  to achieve the diagonal gauge with $(\Lambda^{(+)},\Lambda^{(-)})=(\Lambda,1)$ and $\Lambda$ given in~\eqref{eq:Lor}~\footnote{Notice that it is not possible to set directly  $e^{(+)}= e^{(-)}$ without the help of a Lorentz transformation, as this would be incompatible with the assumption that $\dot e^{(\pm)}$ depend only on $x^\mu$. Therefore we used the gauge $\dot e=\dot e^{(+)}=\dot e^{(-)}$.}. Then the correction is
\begin{align}
\Delta M^{(\text{DFT})}&=b\Delta^{(+)}_{\Lambda} M^{(\text{DFT})}\label{eq:corrM}\\
&+\alpha'U(1+\dot M\Pi)^{-1}\Delta \dot M(1+\Pi\dot M)^{-1}U^T\,,\nonumber
\end{align}
where the second term comes when expanding~\eqref{eq:M} with the $\alpha'$-corrections $\dot M\to \dot M+\alpha'\Delta\dot M$.  Notice that for the heterotic string ($b=0$) the PL map is  uncorrected in the DFT scheme in the gauge $e=e^{(-)}$~\footnote{Alternatively one may choose $e=e^{(+)}$ and the anomalous terms would be $a\Delta^{(-)}_{\Lambda^{-1}} M^{(\text{DFT})}_{nm}$.}.
For the dual background the same reasoning applies, and choosing $\tilde e=\tilde e^{(-)}$
\begin{align}
\Delta \tilde M^{(\text{DFT})}&=b\Delta^{(+)}_{\tilde\Lambda} \tilde M^{(\text{DFT})}\label{eq:corrMt}\\
&\!\!\!\!\!{}+\alpha'\tilde U(\dot MP+\tilde \Pi+\bar P)^{-1}\Delta\dot M(-P\tilde U^{-1}\tilde M+\bar P\tilde U^T)\,.\nonumber
\end{align}

The transformation of the dilatons follows from the fact that the generalized dilaton~\eqref{eq:d} is not anomalous under  Lorentz~\cite{Marques:2015vua} and that the parametrization in terms of standard metric and dilaton holds to $\alpha'$ order. Then
\be\label{eq:corrDiDFT}
\begin{aligned}
\Delta\Phi^{(\text{DFT})}&=\alpha'\Delta\dot d+\tfrac14 G^{mn}\Delta G^{(\text{DFT})}_{mn}\,,\\
\Delta\tilde \Phi^{(\text{DFT})}&=\alpha'\Delta\dot d+\tfrac14\tilde G^{mn}\Delta \tilde G^{(\text{DFT})}_{mn}\,,
\end{aligned}
\ee
where we allowed an $\alpha'$-correction $\dot d\to \dot d+\alpha'\Delta\dot d$. 

We refer to the Supplemental Material for an example of a computation of such $\alpha'$-corrections, and for Refs.~\cite{Eghbali:2018ohx,Eghbali:2020ozg}.
 When specifying the map to a single $U(1)$ T-duality, Eq~\eqref{eq:corrMt} and~\eqref{eq:corrDiDFT} reproduce the rules written in~\cite{Kaloper:1997ux} by Kaloper and Meissner, as proved in~\cite{Borsato:2020bqo}.

\textbf{Conclusions.}
In this Letter we employed the frame-like formulation of DFT to show that (when the conditions in~\eqref{eq:sugra} hold) PL duality is a map between solutions of the low-energy effective string equations at least to first order in $\alpha'$ and {quite possibly} to all orders. We did this for a two-parameter family of theories interpolating between the bosonic and the heterotic string (when the gauge fields and fermions of the latter are set to zero).
It would be interesting to generalize these results to the case in which $G$ for example is replaced by the coset $G/H$.

The importance of Eq.~\eqref{eq:corrM},~\eqref{eq:corrMt} and~\eqref{eq:corrDiDFT} is two-fold. First, they provide the necessary quantum corrections to the PL duality transformation rules in order to extend the map to order $\alpha'$. 
Second, they imply that the form of the $\alpha'$-corrections of backgrounds admitting PL symmetry is strongly constrained by the PL symmetry itself~\footnote{This is under the assumption that the PL symmetry is not anomalous.}. In particular Eq.~\eqref{eq:corrM} and~\eqref{eq:corrDiDFT} can be interpreted as an efficient way to compute $\alpha'$-corrections for PL symmetric backgrounds, since the only unknowns are $\Delta\dot M$ and $\Delta \dot d$, and they can be found by imposing the order $\alpha'$ equations of motion. This is much simpler than trying to compute the corrections directly for $M$ and $\Phi$.
It would be interesting to see if, when considering non-conformal $\sigma$-models, the $\alpha'$-corrections that we find preserve the form of the $\beta$-functions.

\acknowledgements
\section{Acknowledgements}
The work of RB is supported by the fellowship of ``la Caixa Foundation'' (ID 100010434) with code LCF/BQ/PI19/11690019,
by AEI-Spain (FPA2017-84436-P and Unidad de Excelencia Mar\'\i a de Maetzu MDM-2016-0692), by Xunta de Galicia-Conseller\'\i a de Educaci\'on (Centro singular de investigaci\'on  de  Galicia  accreditation  2019-2022, ED431C-2017/07 and ED431G2019/05), and  by FEDER. The work of LW is supported by the grant ``Integrable Deformations'' (GA20-04800S) from the Czech Science Foundation (GACR).

\vspace{12pt}

\textbf{Note added.}
When this work was being written up we learned of the closely related independent work~\cite{Hassler:2020tvz}, and shortly after of~\cite{Codina:2020yma}.

\appendix

\section{Supplemental Material}
In the Supplemental Material we review the essential points of the double formulation, we provide details on the calculation of the generalized fluxes, we review the derivation of the $\alpha'$-corrections, we provide a dictionary to other schemes at order $\alpha'$, and we work out an example.

\textbf{Some details on the double formulation.}
Let us review some basics of the doubled formulation for the reader's convenience.
We work with $O(D,D)$ matrices like
\begin{equation}
\mathcal O_M{}^N=
\left(
\begin{array}{cc}
a^m{}_n & b^{mn}\\
c_{mn} & d_m{}^n
\end{array}
\right),
\end{equation} 
satisfying $\mathcal O_M{}^P\mathcal O_N{}^Q\eta_{PQ}=\eta_{MN}$ where $\eta_{MN}$ was given in~(8). Double curved indices are  lowered and raised with $\eta_{MN}$ and its inverse $\eta^{MN}$. The generalized vielbein $E_A{}^M$ in~(9) can be used to translate curved into flat indices and vice versa, so that
\begin{equation}
\eta_{AB}=
\left(
\begin{array}{cc}
\hat\eta^{ab} & 0\\
0 & -\hat\eta_{ab}
\end{array}
\right),
\end{equation} 
with $ \hat\eta_{ab}$ the Minkowski metric in $D$ dimensions. Double flat indices are  lowered and raised with $\eta_{AB}$ and its inverse $\eta^{AB}$. 
Notice that under a double Lorentz transformation $E_A{}^M\to\Lambda_A{}^BE_B{}^M$ with
\begin{equation}
\Lambda_A{}^B=
\left(
\begin{array}{cc}
\Lambda^{(+)a}{}_b & 0\\
0 & \Lambda_a^{(-)b}
\end{array}
\right)\,,
\end{equation}
the two vielbeins of the standard formulation transform independently under the two copies of the Lorentz group $e^{(+)}\to \Lambda^{(+)}e^{(+)}$, $e^{(-)}\to \Lambda^{(-)}e^{(-)}$.
Another object usually constructed in DFT is the generalized metric $\mathcal H^{MN}$. We do not need it for our derivation, but for completeness let us write its parametrization in curved and flat indices
\begin{equation}
\begin{aligned}
\mathcal H^{MN}&=
\left(
\begin{array}{cc}
	(G-BG^{-1}B)_{mn} & (BG^{-1})_m{}^n\\
	-(G^{-1}B)^m{}_n & G^{mn}
\end{array}
\right),\\
\mathcal H^{AB}&=
\left(
\begin{array}{cc}
	\hat\eta_{ab} &  0\\
	 0 & \hat\eta^{ab}
\end{array}
\right)\,.
\end{aligned}
\end{equation}
At this point let us remark that Eq.~(6) is equivalent to
\begin{align}\label{eq:Mt2}
&\tilde M_{\sigma\tau}=\tilde u_{\sigma i}\tilde N^{ij}\tilde u_{\tau j}\,,\qquad
&&\tilde M_{\sigma \nu}=\tilde u_{\sigma i}\tilde N^{ij}\dot M_{j\nu}\,,\\
&\tilde M_{\mu \tau}=-\dot M_{\mu i}\tilde N^{ij}\tilde u_{\tau j }\,,\qquad
&&\tilde M_{\mu\nu}=\dot M_{\mu\nu}-\dot M_{\mu i}\tilde N^{ij}\dot M_{j\nu}\,, \nonumber
\end{align}
where $\tilde N^{-1}_{ij}=(\dot M+\tilde \Pi)_{ij}$. This way of presenting the transformation rules makes it easier to compare to the known rules of abelian and non-abelian T-duality. Notice that the Lorentz transformation in~(13) can be equivalently written as
\begin{equation}
\tilde \Lambda_a{}^b=\delta_a{}^b-2 \dot e_{ai}\tilde N^{ij}\dot e_j{}^b\,.
\end{equation}

\textbf{Generalized fluxes.}
For the computation of the generalized fluxes it is useful to introduce the generalized Weitzenb\"ock connection
\begin{equation}
\Omega_{ABC}=E_A{}^M\partial_ME_B{}^NE_{CN}\,,
\label{eq:w-conn}
\end{equation}
so that Eq.~(15) becomes
\begin{equation}
\mathcal F_{ABC}=3\Omega_{[ABC]}\,,\quad\mathcal F_A=\Omega^B{}_{BA}+2E_A{}^M\partial_Md\,.
\end{equation}
In order to detail more easily the steps of the computation, let us introduce for convenience $\hat E$ and $\check E$ so that Eq.~(10) is
\begin{equation}
\begin{aligned}
&E=\hat E\mathcal U,\qquad &&\hat E=\dot E(1+\Pi)\,,\\
&\tilde E=\check E\tilde{\mathcal U},\qquad &&\check E=\dot E(1+\tilde\Pi)\,.
\end{aligned}
\end{equation}
Then we can calculate the connection $\Omega_{ABC}$ obtaining
\begin{equation}\label{eq:Omega}
\begin{aligned}
&  \hat E_A{}^\mu\partial_\mu \hat E_B{}^N\hat E_{CN} 
+\hat E_A{}^i \hat E_B{}^J\hat E_C{}^K \mathcal U_i{}^\sigma \partial_\sigma \mathcal U_J{}^N\mathcal U_{KN}\\
&+\hat E_A{}^i\mathcal U_i{}^\sigma\partial_\sigma \hat E_B{}^N\hat E_{CN}\\
& = \hat E_A{}^\mu\partial_\mu \hat E_B{}^N\hat E_{CN} 
+\hat E_A{}^i( \hat E_B{}^j\hat E_{Ck}-\hat E_{Bk}\hat E_C{}^j)T_{ij}{}^k
\\
&+\hat E_A{}^i\mathcal U_i{}^\sigma\partial_\sigma \hat E_B{}^N\hat E_{CN}\\
& = \dot E_A{}^\mu\partial_\mu \dot E_B{}^N\dot E_{CN} 
+\hat E_A{}^i( \hat E_B{}^j\dot E_{Ck}-\dot E_{Bk}\hat E_C{}^j)T_{ij}{}^k
\\
&+\hat E_A{}^i \dot E_{Bj}\dot E_{Ck}\mathcal U_i{}^\sigma\partial_\sigma \Pi^{jk}
\\
& = \dot E_A{}^\mu\partial_\mu \dot E_B{}^N\dot E_{CN} 
+\dot E_A{}^i( \dot E_B{}^j\dot E_{Ck}-\dot E_{Bk}\dot E_C{}^j)T_{ij}{}^k
\\
&+\dot E_{Ai}( \dot E_B{}^j\dot E_{Ck}-\dot E_{Bk}\dot E_C{}^j)\Pi^{il}T_{lj}{}^k\\
&  +\dot E_A{}^i( \dot E_{Bj}\dot E_{Ck}-\dot E_{Bk}\dot E_{Cj})\Pi^{jl}T_{il}{}^k
\\
&+\dot E_{Ai}( \dot E_{Bj}\dot E_{Ck}-\dot E_{Bk}\dot E_{Cj})\Pi^{il}\Pi^{jm}T_{lm}{}^k\\
&
+\dot E_A{}^i \dot E_{Bj}\dot E_{Ck}(\tilde f^{jk}{}_i + 2f_{il}{}^{[j}\Pi^{k]l})
\\
&+\dot E_{Ai} \dot E_{Bj}\dot E_{Ck}\Pi^{il}(\tilde f^{jk}{}_l + 2f_{lm}{}^{[j}\Pi^{k]m})\,.
\end{aligned}
\end{equation}
As shorthand notation we introduced $T_{ij}{}^k = -u_i{}^\sigma u_j{}^\tau \partial_\sigma u_\tau{}^k$, and we used
\begin{equation}\label{eq:dPi}
\partial_\sigma \Pi^{ij} = u_\sigma{}^k(\tilde f^{ij}{}_k + 2 f_{kl}{}^{[i}\Pi^{j]l})\,,
\end{equation}
which can be checked from the definition of $\Pi$.
When we compute $\mathcal F_{ABC}=3\Omega_{[ABC]}$ we find
\begin{equation}
\begin{aligned}
\mathcal F_{ABC}
& = 3\dot E_{[A}{}^\mu\partial_\mu \dot E_B{}^N\dot E_{C]N} \\
&+3\dot E_{[A}{}^i \dot E_B{}^j\dot E_{C]k}f_{ij}{}^k
+3\dot E_{[A}{}^i \dot E_{Bj}\dot E_{C]k}\tilde f^{jk}{}_i\\
&\quad +3\dot E_{[Ai} \dot E_{Bj}\dot E_{C]k}\Pi^{il}(\tilde f^{jk}{}_l + f_{lm}{}^{[j}\Pi^{k]m})\,,
\end{aligned}
\end{equation}
where we used the Maurer-Cartan equation $T_{[ij]}{}^k = \frac12 f_{ij}{}^k$. The last term vanishes due to the identity  $J^{[ijk]}=0$ (see~\cite{Sfetsos:1997pi}) where
\begin{equation}
J^{ijk}=\tilde f^{i[j}{}_l\Pi^{k]l}-f_{lm}{}^i\Pi^{lj}\Pi^{mk}\,.
\end{equation}
Therefore $\mathcal F_{ABC}$ is given by the expression in~(16). As remarked, the computation for $\tilde{\mathcal F}_{ABC}$ is analogous upon exchanging quantities with and without tilde, and lowering or raising indices $i,j,k$.

For the calculation of $\mathcal F_A$ first let us compute
\begin{equation}
\begin{aligned}
\Omega^B{}_{BA}
& = \dot E^{B\mu}\partial_\mu \dot E_B{}^N\dot E_{AN} 
-\dot E_A{}^iT_{ji}{}^j
-\dot E_{Ai}\tilde f^{ij}{}_j\\
&+\dot E_{Ai}\left(-\Pi^{il}T_{jl}{}^j+2f_{jl}{}^{[j}\Pi^{i]l}\right)\,,
\end{aligned}
\end{equation}
where we used that $\Pi$ is antisymmetric. We also have
\begin{equation}
E_A{}^M\partial_Md = \dot E_A{}^\mu\partial_\mu\dot d+\tfrac12 \dot E_A{}^iT_{ij}{}^j+\tfrac12 \dot E_{Ai}\Pi^{il}T_{lj}{}^j\,,
\end{equation}
so that~(17) follows after using again the Maurer-Cartan equation. The analogous computation for the dual background implies
\begin{equation}
\begin{aligned}
\mathcal F_A=\tilde{\mathcal F}_A
&+\dot E_{A}{}^i\left(2f_{ij}{}^j+\tilde f^{jl}{}_{i}\tilde \Pi_{jl}\right) \\
&-\dot E_{Ai}\left(2\tilde f^{ij}{}_j+f_{jl}{}^{i} \Pi^{jl}\right)\,,
\end{aligned}
\end{equation}
so that this flux is invariant if and only if
\begin{equation}\label{eq:sugra0}
2f_{ij}{}^j+\tilde f^{jl}{}_{i}\tilde \Pi_{jl}=0\,,\qquad
2\tilde f^{ij}{}_j+f_{jl}{}^{i} \Pi^{jl}=0\,.
\end{equation}
In fact this condition holds if and only if the structure constants are  traceless as in~(18).
To see this, take for example $g = \exp(\xi^iT_i)$. From the definition~(5) it follows that 
\be
\Pi = \sum_{s=0}^\infty \frac{(-1)^s}{(s+1)!}\Pi_s,\qquad\quad
\Pi_s=\sum_{n=0}^s\binom{s}{n}
(f^T)^{s-n}\tilde f f^{n}\,.
\ee
Here we defined matrices $f,\tilde f$ so that
\be
f_{i}{}^j=\xi^kf_{ki}{}^j,\qquad\quad
\tilde f^{ij}=\xi^k\tilde f^{ij}{}_k\,,
\ee
and $f^T$ is the transpose of $f$.
Obviously $\Pi(\xi=0)=0$, and similarly for $\tilde \Pi$, so that in~\eqref{eq:sugra0} the tracelessness conditions $f_{ij}{}^j=0$ and $\tilde f^{ij}{}_j=0$ are necessary conditions, and we must impose separately $\tilde f^{jl}{}_{i}\tilde \Pi_{jl}=0, f_{jl}{}^{i} \Pi^{jl}=0$. The latter conditions automatically follow from the former, as we are about to prove.
In fact notice that $\Pi_{s+1}=f^T\Pi_s+\Pi_sf$, or in components
\be
\Pi^{ij}_{s+1}=\xi^m(f_{mk}{}^i\Pi_s^{kj}+\Pi^{ik}_sf_{mk}{}^j)\,.
\ee
Then
\be
f_{ij}{}^l\Pi^{ij}_{s+1}=\xi^mf_{mi}{}^l\ f_{jk}{}^i\Pi_s^{jk}\,,
\ee
where we used the Jacobi identity for $f_{ij}{}^k$. Therefore the trace of $f$ with $\Pi_{s+1}$ is proportional to the trace of $f$ with $\Pi_s$, and eventually it is proportional to the trace of $f$ with $\Pi_0=\tilde f$. The same is true for $\Pi$ itself and we find
\be
f_{ij}{}^l\Pi^{ij} = F_k{}^l\ f_{ij}{}^k\tilde f^{ij}{}_m\xi^m\,,\qquad
F=\sum_{s=0}^\infty \frac{(-1)^s}{(s+1)!}(f^T)^s\,.
\ee
The Jacobi identity on the double, apart from the standard Jacobi identities on Lie$(G)$ and Lie$(\tilde G)$, will also impose a quadratic relation between $f_{ij}{}^k$ and $\tilde f^{ij}{}_k$\
\begin{equation}\label{eq:Jac-fft}
4f_{m[i}{}^{[k}\tilde f^{l]m}{}_{j]}-f_{ij}{}^m\tilde f^{kl}{}_m=0\,.
\end{equation}
Contracting indices $i$ and $l$ one finds the identity
\be
f_{ij}{}^k\tilde f^{ij}{}_l = f_{il}{}^k\tilde f^{ij}{}_j+f_{ij}{}^j\tilde f^{ik}{}_l\,.
\ee
We conclude that $f_{ij}{}^l\Pi^{ij}$ vanishes because we are already imposing that the structure constants are traceless. The same reasoning goes for $\tilde f^{jl}{}_{i}\tilde \Pi_{jl}$, and we conclude that the generalized fluxes are invariant if and only if~(18) holds.

\textbf{Anomalous contributions.}
A double Lorentz transformation like $E_A{}^M\to\hat\Lambda_A{}^BE_B{}^M$, with components $\hat\Lambda^{(+)},\hat\Lambda^{(-)}$ receives anomalous contributions at order $\alpha'$, which for $M=G-B$ read~\cite{Marques:2015vua}
\be
\hat\delta M^{(\text{DFT})}_{mn}=\frac{a}{2}\tr(\omega^{(-)}_m\partial_n\hat\lambda^{(-)})+\frac{b}{2}\tr(\partial_m\hat\lambda^{(+)}\omega^{(+)}_n),
\ee
where we omit flat indices and we trace over them, and $\hat\lambda^{(\pm)}$ are the infinitesimal transformation parameters. The finite form of this transformation can be obtained as done in~\cite{Borsato:2020bqo}. Let us set $(\hat\Lambda^{(+)},\hat\Lambda^{(-)})=(\hat\Lambda,1)$ for the sake of the discussion.
Then the finite form of the transformation is
\be\label{eq:finite-Lor}
\begin{aligned}
\hat\Delta_{\hat\Lambda}M^{(\text{DFT})}_{mn}=&-\frac{b}{2}\tr(\partial_m\hat\Lambda^{-1}\hat\Lambda\omega^{(+)}_n)
+bB^{\text{WZW},(\hat\Lambda^{-1})}_{mn}\\
&-\frac{b}{4}\tr(\partial_m\hat\Lambda^{-1}\hat\Lambda\partial_n\hat\Lambda^{-1}\hat\Lambda)
,
\end{aligned}
\ee
where $B^{\text{WZW}}$ was defined in~(22). This formula can be obtained after rewriting
\be
\hat\delta M^{(\text{DFT})}_{mn}=\hat\delta_1 M^{(\text{DFT})}_{mn}+\hat\delta_2 M^{(\text{DFT})}_{mn},
\ee
where
\be
\begin{aligned}
\hat\delta_1 M^{(\text{DFT})}_{mn}&=\frac{b}{2}\tr(\partial_{(m}\hat\lambda\omega^{(+)}_{n)})
+\frac{b}{4}\tr(\partial_{[m}\hat\lambda H_{n]})\\
\hat\delta_2 M^{(\text{DFT})}_{mn}&=\frac{b}{2}\tr(\partial_{[m}\hat\lambda \omega_{n]}),
\end{aligned}
\ee
where we omit flat indices also on $H_{ma}{}^b$. Knowing that under this Lorentz transformation the spin-connection transforms as
\be
\omega\to \hat\Lambda(\omega+d\hat\Lambda^{-1}\hat\Lambda)\hat\Lambda^{-1},
\ee
one notices that the transformation $\hat\delta_1 M^{(\text{DFT})}_{mn}$ agrees with the following infinitesimal transformation
\be
\hat\delta_1 M^{(\text{DFT})}_{mn}=\hat\delta\left[-\frac{b}{4}\tr(\omega^{(+)}_m\omega^{(+)}_n)-\frac{b}{4}\tr(\omega_{[m}H_{n]})\right].
\ee
The finite transformation of the right-hand-side is easy to calculate. For $\hat\delta_2 M^{(\text{DFT})}_{mn}$ one notices that
\be
\hat\delta_2d M^{(\text{DFT})}=\frac{b}{4}\hat\delta CS(\omega),
\ee
where we defined the Chern-Simons form $CS(\omega)=\tr(\omega d\omega+\tfrac23 \omega^3)$ that under Lorentz transforms as
\be
\hat\Delta CS(\omega)=-d\tr(d\hat\Lambda^{-1}\hat\Lambda\omega)-\frac13 \tr[(d\hat\Lambda^{-1}\hat\Lambda)^3].
\ee
Our conventions are such that $B=\tfrac12 dx^n\wedge dx^mB_{mn}$ and $d$ acts from the right. Putting everything together one finds~\eqref{eq:finite-Lor}.
Eq.~\eqref{eq:finite-Lor} gives the anomalous terms when starting from the supergravity frame (i.e. in the diagonal gauge $e^{(+)}=e^{(-)}$) and then doing a Lorentz transformation $e_a^{(+)m}\to \hat\Lambda_a{}^b e_b^{(+)m}$. To obtain the $\alpha'$-corrections to the PL duality map we have to start instead from the situation $e^{(+)}\neq e^{(-)}$ and do the Lorentz transformation $e_a^{(+)m}\to \Lambda_a{}^b e_b^{(+)m}$ where $\Lambda$ was given in~(13). But this can be equivalently obtained by subtracting the corrections coming from the inverse Lorentz transformation. This means that the corrections we are after are
\be
-\hat\Delta_{\hat\Lambda}M^{(\text{DFT})}_{mn},\quad
\text{where }\hat\Lambda=\Lambda^{-1},
\ee
which agrees with~(21). The reasoning to obtain the corrections in the other gauge is analogous.

\textbf{Dictionary to covariant schemes.}
In this Letter, when considering $\alpha'$-corrections to the fields $G_{mn},B_{mn}$ and $\Phi$, we work in a  scheme that is not covariant under local Lorentz transformations, and that is selected when reducing from DFT. In order to translate our results to other covariant schemes typically used in the literature one needs to implement field redefinitions at order $\alpha'$. We collect here the redefinitions to the schemes of Bergshoeff and de Roo (BR)~\cite{Bergshoeff:1988nn,Bergshoeff:1989de}, Metsaev and Tseytlin (MT)~\cite{Metsaev:1987zx} and Hull and Townsend (HT)~\cite{Hull:1987yi}.
From~\cite{Marques:2015vua} we have
\be
\begin{aligned}
M^{(\text{DFT})}_{mn}&=M^{(\text{BR})}_{mn}-\frac{a}{4}\omega^{(-)b}_{ma}\omega^{(-)a}_{nb}-\frac{b}{4}\omega^{(+)b}_{ma}\omega^{(+)a}_{nb}\,,\\
\Phi^{(\text{DFT})}&=\Phi^{(\text{BR})}+\frac{a}{4}\omega^{(-)}_{mab}\omega^{(-)mab}+\frac{b}{4}\omega^{(+)}_{mab}\omega^{(+)mab}\,,
\end{aligned}
\ee
where $\omega_{ma}^{(\pm)b}=\omega_{ma}{}^{b}\pm\frac12H_{ma}{}^{b}$ with $H=dB$,
and the spin-connection is
\be
\omega_{ma}{}^b=e_a{}^n\partial_me_n{}^b-\Gamma_{mn}^pe_a{}^ne_p{}^b\,,
\ee
with $\Gamma_{mn}^p$ the Christoffel symbols. From~\cite{Marques:2015vua} again
\be
\begin{aligned}
M^{(\text{BR})}_{mn}&=M^{(\text{MT})}_{mn}+\tfrac18 (a+b)( H^2_{mn}- 2H_{[m}{}^{ab}\omega_{n]ab}\\
&\qquad\qquad-2\nabla^pH_{mnp}+4H_{mnp}\nabla^p\Phi)\,,\\
\Phi^{(\text{BR})}&=\Phi^{(\text{MT})}+\tfrac{1}{32}(a+b) H^2\,,
\end{aligned}
\ee
where the second line in the redefinition of $M$ vanishes on-shell, and $H^2_{mn}=H_{mpq}H_n{}^{pq}$.
To relate the schemes of HT and MT  we take
\be
\begin{aligned}
&M^{(\text{HT})}_{mn}=M^{(\text{MT})}_{mn}+\tfrac{1}{4}(a+b)H^2_{mn}\,,\\
&\Phi^{(\text{HT})}=\Phi^{(\text{MT})}-\tfrac{1}{16}(a+b)(-1+\tfrac16(1-6q))H^2\,,\\
\end{aligned}
\ee
where the parameter $q$ appears in~\cite{Hull:1987yi}.

\textbf{An example.}
Here we work out explicitly an example to show how to apply our results. We will consider a Bianchi II background that at leading order in $\alpha'$ has metric
\be
ds^2=\frac{2 t (y_1 dy_2-dy_3)^2}{t^2+1}+\frac{\left(t^2+1\right) \left(t\left(dy_2^2+dy_1^2\right)-dt^2\right)}{2 t}\,,
\ee
and $B=0$, $\Phi=0$. We lowered the indices of $y^\sigma$ only for readability. This background is Ricci flat and therefore solves the supergravity/one-loop equations. Adding enough flat directions one gets to $D=26$.

This metric is invariant under the $G$ isometry group generated by $T_i, i=1,2,3$ with
\be
[T_1,T_2]=T_3\,.
\ee
We will apply NATD on this background, so that we have $\tilde G$ abelian $\tilde f^{ij}{}_k=0$.
This also implies $\Pi=0$. This background is of the form~(4) if we take $g=\exp(y^2T_2)\exp(y^1T_1)\exp(y^3T_3)$ and $\dot M$ diagonal with $\dot M_{11}=\dot M_{22}=t\dot M_{33}^{-1}=-t\dot M_{44}=\tfrac12 (1+t^2)$. Here we are enumerating coordinates as $x^m=(y^1,y^2,y^3,t)$ and we will dualize all coordinates $y$.
Moreover from~(14) we have $\dot d=-\tfrac14 \log(\tfrac14(1+t^2)^2)$.

It is straightforward to obtain the dual background at leading order in $\alpha'$. Since $\tilde G$ is abelian, $\tilde U=1$ and we have the only non-vanishing components $\tilde \Pi_{12}=-\tilde \Pi_{21}=\tilde y_3$.
Then the dual metric is diagonal with $\tilde G_{11}=\tilde G_{22}=2(1+t^2)(4\tilde y_3^2+(1+t^2)^2)^{-1}$, $\tilde G_{33}=\tfrac12 t^{-1}(1+t^2)$, $\tilde G_{44}=-\tfrac12 t^{-1}(1+t^2)$, the $B$-field is just $\tilde B_{12}=-\tilde B_{21}=4\tilde y_3(4\tilde y_3^2+(1+t^2)^2)^{-1}$ and the dilaton $\tilde \Phi=\dot d+\tfrac14 \log[(1+t^2)^4(t+4\tilde y_3^2t+2t^3+t^5)^{-2}]$.

We now want to obtain $\alpha'$-corrections for the original and dual backgrounds, so that the two-loop equations of the bosonic string are satisfied. We will work in the scheme of Hull and Townsend (HT) at $q=0$~\cite{Hull:1987yi}. The relevant two-loop equations in our conventions can be read from~\cite{Borsato:2019oip}. In~\cite{Eghbali:2018ohx,Eghbali:2020ozg} it was noticed that in certain examples one can apply the usual rules of NATD without further corrections, and still be able to obtain two-loop solutions. In general further corrections to the rules are necessary, as we are about to see in this example.
There are three types of contributions to take into account: (i) the anomalous transformations in the first line of~(23), (ii) the second line of~(23) coming from expanding~(4) at next order, (iii) the redefinitions to map the DFT scheme to the HT one.

First we have to pick a choice for the vielbeins. Since $\dot M$ is diagonal, it is natural to choose also $\dot e$ diagonal. Since $\Pi=0$, the Lorentz transformation $\Lambda=1$ and then we immediately have $e=e^{(+)}=e^{(-)}=U\dot e$, and in this case there will be no contributions of type (i). We need this vielbein also to compute the contributions (iii) since 
\be
M^{(\text{HT})}_{mn}=M^{(\text{DFT})}_{mn}-\tfrac{1}{2}\alpha'\omega^{(-)b}_{ma}\omega^{(+)a}_{nb}\,.
\ee
In the contributions (ii) $\Delta \dot M$ is an unkown to be fixed by imposing the two-loop equations. The same is true for $\Delta\dot d$ which appears in the correction to the dilaton. In fact the correction to the dilaton in the HT scheme can be obtained by noticing that~(25) implies
\be
\Delta\Phi^{(\text{HT})}=\alpha'\Delta\dot d+\frac14 G^{mn}\Delta G^{(\text{HT})}_{mn}+\frac{\alpha'}{48}H^2\,,
\ee
where $\Delta G^{(\text{HT})}_{mn}$ is the full correction to the metric taking into account (i),(ii) and (iii). We find that the original background solves the two-loop equations if we consider all these $\alpha'$-corrections and if we set for example $\Delta\dot d=[t \left(-2 t^4+3 t^2+2 \left(-t^4+t^2+2\right) t \arctan t+3\right)]/\gamma$ with $\gamma=\left(t^2-1\right)^{2} \left(t^2+1\right)$ and if we take $\Delta\dot M$ diagonal with $\Delta\dot M_{11}=\Delta\dot M_{22}=\tfrac12[\left(3 t^2-1\right) \left(t^2+1\right)^3 \arctan t+t \left(3 t^6+7 t^4+9
   t^2-3\right)] \left(t^2-1\right)^{-1} \left(t^2+1\right)^{-2}
$, $\Delta\dot M_{33}=(t^4-6 t^2+1)\left(t^2+1\right)^{-4}$ and $\Delta\dot M_{44}=[-2 t^8+20 t^4+8 t^2-2 \left(t^2-3\right) \left(t^2+1\right)^3 t \arctan t+6]\left(t^4-1\right)^{-2}$.

To get the $\alpha'$-correction to the dual background we apply the same procedure. The difference now is that $\tilde e^{(+)}\neq \tilde e^{(-)}$ and therefore $\tilde \Lambda$ is non-trivial. Choosing $\tilde e=\tilde e^{(-)}$ we can obtain the corrections of type (i) as in the first line of~(24). In this example the WZW-like contribution to the $B$-field turns out to be zero. Obviously in~(21) we have to use the dual vielbein $\tilde e$ and spin-connection $\tilde \omega$.
The same is true when considering corrections (iii). Putting together all corrections (whose explicit form we omit for brevity) of type (i), (ii), (iii) and of the dilaton, and using the previous $\Delta\dot M$, $\Delta\dot d$, one can check that indeed the $\alpha'$-corrected dual background correctly solves the two-loop equations. Obviously, working with the other choice $\tilde e=\tilde e^{(+)}$ gives the same result since we are in the covariant HT scheme, provided the anomalous transformations are now obtained from $a\Delta^{(-)}_{\tilde\Lambda^{-1}}M^{(\text{DFT})}_{nm}$. We remind that in the bosonic string we have to set $a=b=-\alpha'$.

\bibliographystyle{h-physrev}
\bibliography{
biblio}

\end{document}